\documentclass[usegraphicx,usenatbib,useapjfonts,apj]{emulateapj}
\usepackage{graphicx}  
\usepackage{dcolumn}   
\usepackage{bm}        
\usepackage{amssymb}   

\def\be{\begin{equation}}
  \def\ee{\end{equation}}
\def\bea{\begin{eqnarray}}
  \def\eea{\end{eqnarray}}
\def\fun#1#2{\lower3.6pt\vbox{\baselineskip0pt\lineskip.9pt
\ialign{$\mathsurround=0pt#1\hfil##\hfil$\crcr#2\crcr\sim\crcr}}}

\shorttitle{Galactic kSZ}
\shortauthors{Hajian et al.}

\begin{document}

\title{The Kinetic Sunyaev-Zel'dovich Effect Due to the Electrons of Our Galaxy}
\author{Amir Hajian\altaffilmark{1,2}, Carlos Hern\'andez-Monteagudo\altaffilmark{3}, Raul Jimenez\altaffilmark{3}, David Spergel\altaffilmark{2}, Licia Verde\altaffilmark{3}} 
\altaffiltext{1}{Department of Physics, Jadwin Hall, Princeton University,
    Princeton, NJ 08542; {\it ahajian@princeton.edu}.}
\altaffiltext{2}{Department of Astrophysical Sciences, Peyton Hall,
   Princeton University,
    Princeton, NJ 08544; {\it dns@astro.Princeton.edu}. }
\altaffiltext{3}{Department of  Physics and Astronomy, University of Pennsylvania, 209s 33rd str, Philadelphia, PA 19104; {\it carloshm@astro.upenn.edu; 
raulj@physics.upenn.edu;lverde@physics.upenn.edu}.}

\begin{abstract}
We compute the effect of local
electrons on the CMB temperature anisotropies.
The number density and distribution of free electrons in our Galaxy
has been accurately measured from pulsar dispersion measurements.
Because of their distribution, the dynamics of our Galaxy and the
Galaxy peculiar velocity with respect to the Hubble flow, these free
electrons leave a frequency-independent imprint on the cosmic
microwave background (CMB). In particular, the coherent motion of the
free electrons respect to us and to the CMB rest frame produce a
kinetic Sunyaev-Zeldovich signal. We compute this effect and 
we note that the large-scale antisymmetry of the signal gives 
it an angular power spectrum with a sawtooth pattern where 
even multipoles are suppressed with respect to the odd ones.  
We find the signal to
be small ($\sim 2 \mu K$) and sub-dominant compared to the primary CMB
and other foreground signals. However, since there are no free
parameters in the modeling of this signal, it can be taken into
account if more precise measurements of the primordial signal are
required.

\end{abstract}


\section{Introduction and Background}
The motion of the ionized gas in our Galaxy leaves its imprint on the CMB 
through the Kinetic Sunyaev-Zeldovich Effect \citep{SZ1, Hogan}. 
These distortions are spectrally indistinguishable from the CMB. 
 Thompson scattering of
CMB photons from a line element $ds$ of free electrons with optical
depth $d\tau$ along the line of sight determined by the unit vector
$\hat{\bf n}$, gives a correction to the CMB temperature:
\begin{equation}
dT(\hat{\bf n})=-d\tau \delta T_{CMB}(\hat{\bf n}) - d\tau \frac{\bf v}{c}\cdot \hat{\bf n} T_{CMB}(\hat{\bf n}).
\end{equation}
Thus there are two terms, the first is a blurring of the CMB
anisotropies ($\delta T_{CMB}(\hat{\bf n})=T_{CMB}(\hat{\bf n})-T_0$ where $T_0=\langle T_{CMB}\rangle$), while the
second is a generation of new anisotropies: the kinetic SZ (kSZ) effect.

If $\delta T_{CMB}\ll T_{0}$ (i.e., we are in a coordinate system where the
CMB has zero dipole\footnote{Note that this will not be the case in a
coordinate system where the CMB has a large dipole, like for example
in a coordinate system where the Sun is at rest}) and $v/c > 10^{-5}$, the second term dominates. Thus the integrated effect along the line of sight, for $\tau \ll 1$ is given by:
\begin{equation}
\label{eqn:kszint}
\frac{\delta T}{T}=-\int n_e \sigma_T \frac{v_r}{c} dl \sim \tau(\hat{\bf n})\frac{v_r(\hat{\bf n})}{c}
\end{equation}
where $v_r \equiv {\bf v}\cdot \hat{\bf n}$,  $n_e$ denotes the electron density and $\sigma_T$ the Thompson
cross section.  Given that plausible velocities are of the orders of
few $\times 100$ km/s (the largest velocity, $620$ km/s, is that of
the infall of the local group to the Great Attractor) only  $\tau > 10^{-3}$ yields a measurable signal. Such high optical depth
implies a number density of electrons and baryonic mass that is only
consistent with that present in the disk of our own Galaxy.

In this case, the velocity can be written as a sum of the Sun's
velocity as measured from the CMB dipole, $v_{\rm dipole}$, and the gas velocity 
with respect to the Sun.
Note that in a coordinate system where  the Sun is at rest, the  contribution for $v_{\rm dipole}$ is interpreted as a blurring of anisotropies (first term in the LHS of eq 1), the anisotropies being the  CMB dipole.

Here, we compute the kSZ signature of the Galaxy on the CMB
sky. While this signal is sub-dominant compared to other foreground
signals and small compared to the intrinsic CMB anisotropy, it is a
component that must be there and that can be modeled as we illustrate
below.  We assume a simple model for the velocities of the free
electrons in our Galaxy that includes an axisymmetric model for the
rotation around the Galactic center with the same speed as the
luminous matter, $\mathbf{v}_{rot}$, and a bulk motion,
$\mathbf{v}_{bulk}$ of the galactic center with respect to the Hubble flow.  
We neglect the higher order corrections to these
assumptions  such as blurring of intrinsic CMB anisotropies (first term in eq. 1). 
The total velocity of the free electrons with respect to
the CMB rest frame is the vector sum of the above velocities,
$\mathbf{v}=\mathbf{v}_{rot}+\mathbf{v}_{bulk}$. The bulk velocity,
$\mathbf{v}_{bulk}$, can be derived from the observed dipole of the
CMB as $\mathbf{v}_{bulk}=\mathbf{v}_{dipole}-\mathbf{v}_\odot$, where
$\mathbf{v}_\odot$ is the linear velocity of the rotation of the Sun
around the galactic center. Therefore we have
$\mathbf{v}=\mathbf{v}_{dipole}+(\mathbf{v}_{rot} - \mathbf{v}_\odot)$. 
The radial part of this velocity is the $v_r$
that appears in eqn.(\ref{eqn:kszint}) and is given by 
\be
\label{eqn:vr}
v_r=v_{los}+\mathbf{v}_{dipole}\cdot\hat{\mathbf{n}}, 
\ee 
where $v_{los}=(\mathbf{v}_{rot} - \mathbf{v}_\odot)\cdot\hat{\mathbf{n}}$
is the line-of-sight velocity with respect to the Sun.  


\section{The Model for the Galactic Distribution of Free Electrons}
We use the Cordes and Lazio \citep{Cordes:2002wz} model (hereafter NE2001) 
 for the distribution of free electrons in our Galaxy. 
\cite{Cordes:2002wz} combine the measurements of the 
dispersion measures (DM) of pulsars,
temporal broadening of pulses from pulsars with large DM,
scintillation bandwidth measurements of low-DM pulsars, angular
broadening of Galactic and extragalactic sources and emission
measures  to infer the distribution of the free electrons in the Galaxy
responsible for pulsar dispersion measures and a spatial model
of the warm ionized component of the interstellar gas. 
\begin{figure}[h]
\plotone{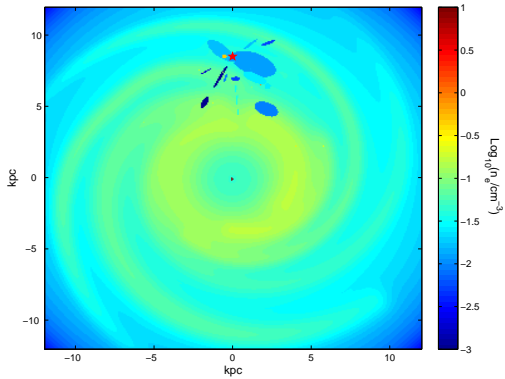}
\caption{\label{fig:ne} Distribution of the free electrons in our
Galaxy, $log_{10}(n_e)$, based on NE2001. The $n_e$ is shown at the
Galactic plane, $z=0$, and is maximum at the Galactic center. The
position of the Sun is denoted by a star, $\star$, and the small
overdense region close to the Sun is the Gum Nebula.
}
\end{figure}

The electron density distribution is given by the sum of two
axisymmetric components (the thick disk and the thin disk) and a
spiral arm component, combined with terms that describe specific
regions in the Galaxy. The local interstellar medium
is modeled with multiple components: 
(a) four low density regions near the Sun: a local hot
bubble centered on the Sun's location, the North Polar Spur, a local
superbubble and another low density region, (b) the Galactic center
component, (c) the Gum Nebula and  the Vela Supernova Remnant, (d) regions of
intense scattering (clumps) and (e) regions of low density (voids).

Figure \ref{fig:ne} shows the Galactic distribution of the free
electrons, $n_e$,at the Galactic plane,  based on the NE2001 model. The position of the Sun is denoted by a red star. The high electron density spot on the left side of the Sun is the  Gum nebula. 

\section{Kinematics of the Differential Rotation of the Galaxy}

We consider an axisymmetric model for the rotation of the Milky
Way. The orbits in this model are circular and the angular velocity at
point $\mathbf{R}$ is $\mathbf{\Omega}(R)$, where $R=|\mathbf{R}|$. We
are interested in the line-of-sight velocity, $v_{los}$, the
projection of the velocity of each particle relative to the Sun along the vector connecting the Sun to
that point, $({\mathbf R} - {\mathbf R}_{\odot})$. Here, ${\mathbf R}_{\odot}$
denotes the position of the Sun in the Galactocenteric coordinates.
Therefore, 
\be
v_{los}=\hat{{\mathbf R}}_{los}\cdot ({\mathbf v}_{rot}-{\mathbf v}_{\odot}),
\ee 
where $\hat{{\mathbf R}}_{los}=({\mathbf R} - {\mathbf R}_{\odot})/|{\mathbf R} - {\mathbf R}_{\odot}|$.
Following the standard discussion in  \cite{GalacticAstronomy}, the  line of sight velocity $v_{los}$  can be expressed as a function of longitude, $l$, 
on the celestial sphere and $R$
\be
\label{vlos}
v_{los}(l,R) = \left[ \Omega(R) - \Omega(R_{\odot})\right] R_{\odot} \sin{l}.
\ee

The disk of the Milky Way consists of three parts: the central disk at
$R<3$ kpc, the inner disk, $R_0>R>3$ kpc and the outer disk,
$R>R_0$. The determination of the $v_{los}$ is different in each of
the three parts.

For the inner disk, the observable is the terminal velocity,
$v_{los}^t(l_1)$, along each line of sight, $l_1$. The terminal
velocity is defined as follows: consider the smallest ring -centered
on the galactic center- that intercepts the line of sight. The line of
sight is thus tangent to this ring, and the radius of the ring is $R = R_{\odot} \sin{l_1}$. The terminal velocity is the line of sight velocity of the ring at the tangent point. 21-cm line and  CO emission line observations 
are used to estimate  $v_{los}$ (e.g., \cite{rougoor}) 
and  the distance to the tangent point is determined by geometry.
The circular velocity of the ring can be calculated using $v_{los}^t(l)$ values reported in Chapter 9 of \citet{GalacticAstronomy} and eq.(\ref{vlos}),
\be 
v_c(R) =
v_{los}^{t} (l_1) + v_c(R_0) \sin{l_1}, \,\,\,\,\,\,\sin{l_1}=R/R_0,
\ee 
and therefore 
\be 
v_{los}(l,R) = \left[v_c(R) \cdot R_0/R -v_c(R_0)\right] \sin{l}.  
\ee

In the outer disk, distances  are not so easily determined and thus distances and  velocities are obtained either by 
observations of Cepheid variables or  by main sequence fitting of a 
young cluster and radio-line observations of  associated molecular 
gas (see \citet{bb93} and references therein).

The measurement of $v_{los}$ is then given by 
\be
v_{los}(l,R) =  W(R) \sin{l}, 
\ee
where $W(R) \equiv R_0[\Omega(R)-\Omega(R_0)]$ is given as a function of $R/R_0$ in \citet{bb93}. Note however that the errors associated to the distance $R$ and the errors assigned to a given measurement of $W(R)$ increase dramatically for $R>1.5R_0$, and therefore very little can be said about $v_{los}$ for radii larger than 2$R_0$, \citep{binneydehnen97}.

\begin{figure}[h]
  \plotone{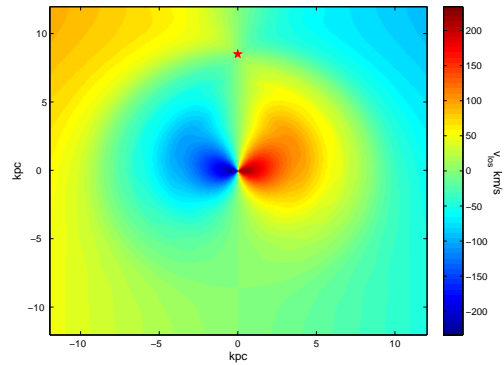}
  \caption{\label{fig:vlos} The line-of-sight velocity relative to the Sun, $v_{los}$, on the plane of the Galaxy.The position of the Sun is denoted by $\star$.   $v_{los}$ is negative when a point moves towards the Sun.  }
\end{figure}

The velocities in the central disk are more complicated. Here, we
calculate them in the same way as the inner disk velocities: we
extrapolate the terminal velocities of the central disk from those of
the inner disk. This approximation however does not significantly affect our final
result because the $v_{los}$ is small in the vicinity of the Galactic
center. In addition, the effect of non-circular motions such as
axisymmetric expansion, oval distortions of the orbits and random
motions, act as a correction to the above model. For
simplicity we ignore these corrections and only consider circular
motions: as discussed in Chapter 9 of \citet{GalacticAstronomy}, this simplified model works well outside of the central (3$kpc$) disk.

\section{The kinetic SZ Pattern}
We now separately compute the contribution to the kSZ signal due to the 
rotation of the Galaxy (\ref{sub1}) and the contribution due to the motion of the Galaxy with respect to the CMB rest frame (\ref{sub2}).
 We use the free electron density model of NE2001 and the above model
for the velocity field of the Galaxy to obtain the Galaxy kSZ pattern.
As the velocity field, $v_{los}$ is defined with respect to the Sun, there will be an additional contribution due to ${\bf v}_{dipole}$. 

\subsection{Motion within the Galaxy}
\label{sub1}
\begin{figure}[h]
  \includegraphics[width=0.3\textwidth, angle=90]{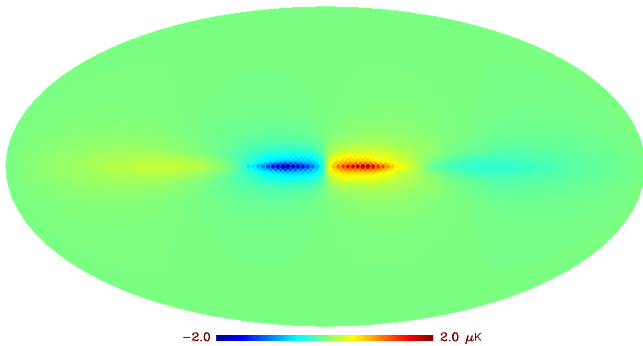}
  \caption{\label{fig:ksz} Map of the predicted kinetic SZ signal
  caused by the rotation of the free electrons in the disk of
  our Galaxy.}
\end{figure}

We assume the Sun is located at $(x,y,z) = (0.0,
8.5, 0.0)$ kpc in Galactocenteric coordinates,
$R_0=8.5$ kpc, and  $v_{\odot}=220$ km/s. The kSZ effect is calculated using eq.(\ref{eqn:kszint}). 
We integrate  eq.(\ref{eqn:kszint}) along the line of sight: 
\bea
\label{eqn:kszdiscrete}
\Delta T(\hat{n}) &=& -  \frac{\sigma_T}{c}  T_0 \sum_{i=1}^{N} n_e(r_i \hat{n})v_{los}(r_i \hat{n}) \Delta r\\ \nonumber
&\simeq&- (4.1 \mu K)
\sum_{i=1}^{N} \left(\frac{n_e(r_i \hat{n})}{1 cm^{-3}}\right) \left(\frac{v_{los}(r_i \hat{n})}{220 {\rm km/s}}\right)  \left(\frac{\Delta r}{kpc}\right),
\eea
where $r_i=r_{i-1}+\Delta r$ with $r_0=0$.  
While the velocities in the outer disk are not known beyond $r\sim 1.5-2
R_0$, most of the contribution comes from the central and inner disks, hosting the largest electron densities and peculiar velocities. Thus  we only consider electrons that are within the $r=2R_0$ sphere. 
The result is shown in Fig. \ref{fig:ksz}.
 The kSZ pattern is  antisymmetric: its main
features are a cold and a hot spot at
the two sides of the Galactic center with $|l|<90^\circ$ and
$\delta T_{max}\sim 2\mu K$. These spots come from the regions in the inner
disk with the highest $v_{los}$ (see Fig. \ref{fig:vlos}).  There are
weaker warm and cold spots at larger longitudes, $|l|>90^\circ$,
caused by the rotating electrons in the outer disk. The kSZ signal is
zero at $l=0^\circ$ and $l=180^\circ$ where $v_{los}=0$. Both the
Galactic center and the Gum Nebula, which have the larger $n_e$, happen
to fall in regions with very small $v_{los}$ and therefore do not
contribute much to the kSZ  due to the rotation of the Galaxy.

\subsection{Motion of the Galaxy}
\label{sub2}
The second term of 
eq.(\ref{eqn:vr}) arises from the bulk motion of the Galaxy with 
respect to the CMB rest frame.
We consider the dipole velocity to be
$v_{dipole}=371km/s$ in the direction of 
$(l,b)=(264^\circ,48^\circ)$.
\begin{figure}[h]
  \includegraphics[width=0.3\textwidth, angle=90]{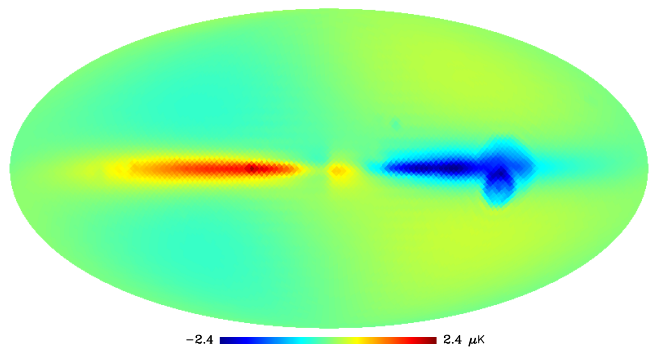}
  \caption{\label{fig:final} Dipole subtracted map of the effect of the free electrons in our Galaxy due to the rotation and the bulk motion of the 
Milky Way.}
\end{figure}
This bulk motion generates a kSZ signal bigger than the one due to the
rotation of the Galaxy. The signal can be interpreted as a suppression of the dipole due to the free electrons in the Galaxy, thus it is a  negative dipole modulated by the line of sight integration of the electron distribution $n_e$.
The signal of the  Galactic Center and the Gum nebula feature prominently because of their  high density. 

\subsection{The Combined effect}
\begin{figure}[h]
  \includegraphics[width=0.4\textwidth, height=2in, angle=0]{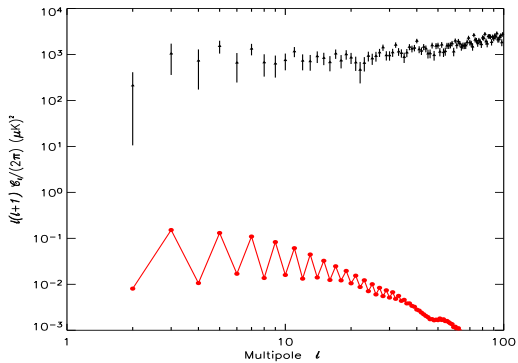}
  \caption{\label{fig:kszcl} Angular power spectrum $C_{l}$ for the  predicted Kinetic SZ signal (red filled circles). Measurements from WMAP (3yr data) are displayed as black triangles.}
\end{figure}

 We add this to the map of Figure \ref{fig:ksz}
to obtain the total kSZ signal generated by the free electron
distribution in our Galaxy. The resulting  map {\em after dipole subtraction}
is shown in Figure \ref{fig:final}. This pattern is still antisymmetric
and mostly lies in the plane of the Galaxy but it 
overrides the four-fold pattern of Figure \ref{fig:ksz}.
The cold spot centered at $(l,b)=(-90^\circ,0^\circ)$ is
due to the overdensity of the free electrons in the Gum Nebula and the
warm spot at the center of the Galaxy  reflects the high $n_e$ there.
Figure \ref{fig:kszcl} shows the angular
power spectrum  of the kSZ signal. Most of the power resides at large scales
(low $l$'s), but its contribution to the intrinsic CMB power spectrum
remains around the 0.01\% level.
An interesting feature of the above pattern is the saw-tooth shape
of its  angular power spectrum. This is due to the
large-scale antisymmetry of the kSZ pattern which suppresses the even
multipoles with respect to the odd ones.  The pattern quickly
disappears at higher multipoles.

\section{Discussion and Conclusion}
We have computed the kinetic SZ contribution to the CMB anisotropies
due to the coherent motion of the free electrons in the disk of our
Galaxy. We used the model of \cite{Cordes:2002wz} for the galactic
distribution of free electron and kinematics of the differential
rotation of our Galaxy to compute the line of sight velocity with
respect to the Sun. The bulk motion of our Galaxy causes another effect
that is comparable to this effect. We calculate the anisotropic pattern due to 
the combination  of these effects and show the result in Fig. \ref{fig:final}.

This kSZ signal  is subdominant compared to the
primary CMB and other foregrounds, but the model used to compute it has
no free parameters and thus this contribution to the anisotropy can be
easily modeled and subtracted out. This effect has been independently 
studied by another group \citep{Waelkens} and our results agree. The polarization signal of the Thompson scattering from the free electrons in the local Universe was studied in \cite{Hirata:2005nw}.
\begin{figure}[h]
  \plotone{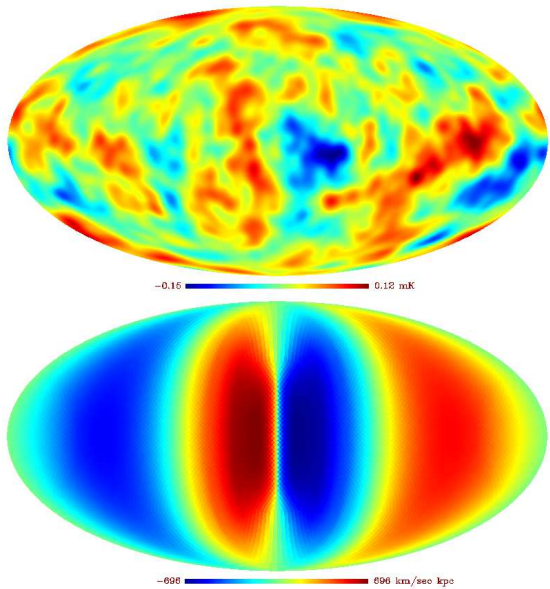}
  \caption{\label{fig:vlos_map} {\emph{top:}} The ILC map based on the WMAP 3yr data, smoothed with a $420'$ beam.  {\emph{bottom:}} A map of the integral of the line-of-sight velocity relative to the Sun in Galactic coordinates. Both maps happen to have  a similar structure on the large scales.} 
\end{figure}
In Figure \ref{fig:vlos_map} we show WMAP's  ILC map \cite{Hinshaw:2006ia} smoothed with a $420'$ beam and a map in galactic coordinates of the integral of the line of sight velocity due to the rotation of the Galaxy. Note the  similarity between the maps.  
However, any scattering process would flip the sign of the radial velocities.
We find another similarity in the saw-tooth pattern of the angular
power spectrum of the kSZ map (Fig \ref{fig:kszcl}). This pattern has
the same shape as the full-sky power spectrum derived from WMAP's ILC
map: at low-$l$, even multipoles are suppressed with respect to the
odd ones. Although the kSZ signal is too small to explain the
suppression of the even multipoles in the WMAP data, it suggests that
a mechanism that boosts the above kSZ signal by about two orders of
magnitude might explain the saw-tooth pattern of the ILC power
spectrum and other peculiarities of the two point correlation of the
ILC map (such as the anticorrelation at $\theta=180^\circ$,
\cite{Hajian_in_prep,Hajian_ctheta}).  
Any element in our Galaxy that does not emit
but only scatters CMB photons would be a good candidate for the above
process, since it would preserve the thermal CMB spectrum and enhance
the kSZ signal. We do not know any physical mechanism with the above properties.The intriguing similarity in figure  \ref{fig:vlos_map} is, therefore more likely a warning about the dangers of  {\it a posteriori} analysis  than the signature of novel physics.
\acknowledgements 
AH thanks Joseph Taylor and  Andre Waelkens for enlightening discussions.  Some of the results in this paper have been derived using the HEALPix package \citep{Gorski:2004by}. 
We acknowledge the use of the Legacy Archive for Microwave Background Data Analysis (LAMBDA). Support for LAMBDA
is provided by the NASA Office of Space Science. AH acknowledges 
support from NASA grant LTSA03-0000-0090. LV and CHM are supported
in part by NASA grant ADP03-0000-009 and ADP04-0000-093. RJ, LV, CHM
and DNS are supported in part by NSF PIRE-0507768 grant.  LV, RJ and
CHM thank the Physics department of Princeton University for
hospitality while part of this work was being carried out.

\end{document}